\documentclass[aps,prl,twocolumn,amsmath,amssymb]{revtex4-2}

\usepackage{mathrsfs}
\usepackage{graphicx}
\usepackage{bm}
\usepackage{threeparttable}
\usepackage{amsmath,amssymb}
\usepackage{color}
\usepackage{epstopdf}
\usepackage[colorlinks=true,linkcolor=blue,anchorcolor=blue, citecolor=blue,urlcolor=blue]{hyperref}
\usepackage{float}

\begin{document}

\preprint{APS/123-QED}

\title{Synchronization and phase transition of two-dimensional self-rotating clock models}

\author{Xin Wu}
\email{wuxin193@mails.ucas.ac.cn}
\affiliation{Beijing National Laboratory for Condensed Matter Physics and Laboratory of Soft Matter Physics, Institute of Physics, Chinese Academy of Sciences, Beijing 100190, China}\affiliation{School of Physical Sciences, University of Chinese Academy of Sciences, Beijing 100049, China}

\author{Mingcheng Yang}
\email{mcyang@iphy.ac.cn}
\affiliation{Beijing National Laboratory for Condensed Matter Physics and Laboratory of Soft Matter Physics, Institute of Physics, Chinese Academy of Sciences, Beijing 100190, China}\affiliation{School of Physical Sciences, University of Chinese Academy of Sciences, Beijing 100049, China}

\begin{abstract}
We explore possible synchronization in two-dimensional (2D) locally coupled discrete-state oscillators under thermal fluctuations, using the self-rotating $q$-state clock model as a prototype. Large-scale Monte Carlo simulations reveal that for $q \ge q_c$ (with $q_c = 5$), the system undergoes two-step Berezinskii-Kosterlitz-Thouless (BKT)-like transitions: first from a disordered phase to a critical synchronized phase, and then to a spatiotemporal pattern phase. Notably, the synchronized phase features algebraically decaying spatial correlations and divergent coherence time, realizing an effective continuous time crystal across macroscopic yet finite scales; while it vanishes when $q < q_c$. A dynamic renormalization group analysis shows this behavior arises from an emergent U(1) symmetry for $q \ge q^{RG}_c=5$, and indicates a crossover scale to Kardar-Parisi-Zhang (KPZ) universality diverges double-exponentially with $q$, ensuring the pre-asymptotic stability of the synchronized phase. Mean-field theory predicts a lower critical value $q_c^{MF} = 4$.
\end{abstract}

\maketitle

\emph{Introduction.}---
Synchronization is a longstanding fundamental topic in statistical physics and nonlinear science. Phase synchronization of continuous-state coupled oscillators has been extensively studied, with the Kuramoto model as a paradigm~\cite{Kuramoto}—whereas far less attention has been devoted to the synchronization of discrete-state oscillators under thermal fluctuations.
Mathematically, it has been proved that synchronization can occur in noisy discrete-state oscillator systems with 3D short-range interactions~\cite{nonergodic} and with 1D or 2D long-range interactions~\cite{onetwodimensional}. However, whether synchronization emerges among noisy discrete-state oscillators with 2D short-range interactions remains unresolved. 

Meanwhile, in statistical physics, more and more discrete-state dynamic models have been developed and explored due to their simplicity and universality. Prominent examples include: the unidirectionally rotating 3-state oscillator model~\cite{wood_PhysRevE,wood_PhysRevLett,rosas_synchronization_2020}; the non-reciprocal Ising model (equivalent to a rotating 4-state clock model)~\cite{Nonreciprocal_pre,Nonreciprocal_prl,Guislain_2024,twopopulationIsing}; the dissipative Ising model~\cite{dai_pra_curie,rhythmicisingmodel}; and the driven or active Potts model~\cite{MinimalNPT,driven_Potts,syndriven_Potts,ScalingCorrelations,noguchi_cycling_2024,noguchi_spatiotemporal_2025,Noguchi_2025}. These studies apparently suggest that synchronization of locally coupled discrete-state oscillators appears only in dimensions three and above and is absent in two dimensions~\cite{swart_course,wood_PhysRevE,Nonreciprocal_pre}.

In contrast to the above observations, even the simplest noisy continuous-state oscillators with 2D short-range couplings, i.e., the minimal Kuramoto model or infinite-state rotating clock model, can exhibit synchronization~\cite{sarkar_noiseinduced_2020,Self_Driven_Rotors}. This sharp discrepancy between the $q$-state oscillator systems with $q\le4$ and $q=\infty$ raises a fundamental question: how and why does the value of $q$ shape the synchronization and phase behavior? To seek instructive insights, we revisit the well-established equilibrium 2D $q$-state clock model~\cite{Renormalization_planar,discrete_spin,clock_2019,clock_2020,clock_2022,clock_2022_prr}. When $q\ge 5$, the equilibrium clock model displays three distinct phases—disordered, quasi-long-range ordered, and long-range ordered—separated by two-step BKT transitions at high and low temperatures; while for $q < 5$, the quasi-long-range ordered phase is absent. Inspired by these contrasts, we propose a natural hypothesis: the absence of synchronized phases in existing noisy discrete-state oscillator models with 2D short-range interactions may stem from the lack of a quasi-long-range ordered phase in their equilibrium counterparts.

In this Letter, we explore the above hypothesis by systematically studying the phase behavior of 2D locally coupled self-rotating $q$-state clock model—a system whose equilibrium counterpart and continuous version are well understood~\cite{oscillatorsdiscrete,discreteKuramoto}. We find that the system exhibits a rich variety of $q$-dependent phases and transitions. Specifically, for $q \ge 5$, a critical synchronized phase emerges with quasi-long-range order and time crystal behavior, whose asymptotic destruction by KPZ nonlinearity is suppressed by a double-exponentially diverging crossover scale.

\begin{figure}[htbp!]
    \centering
    \includegraphics[keepaspectratio, width=\columnwidth]{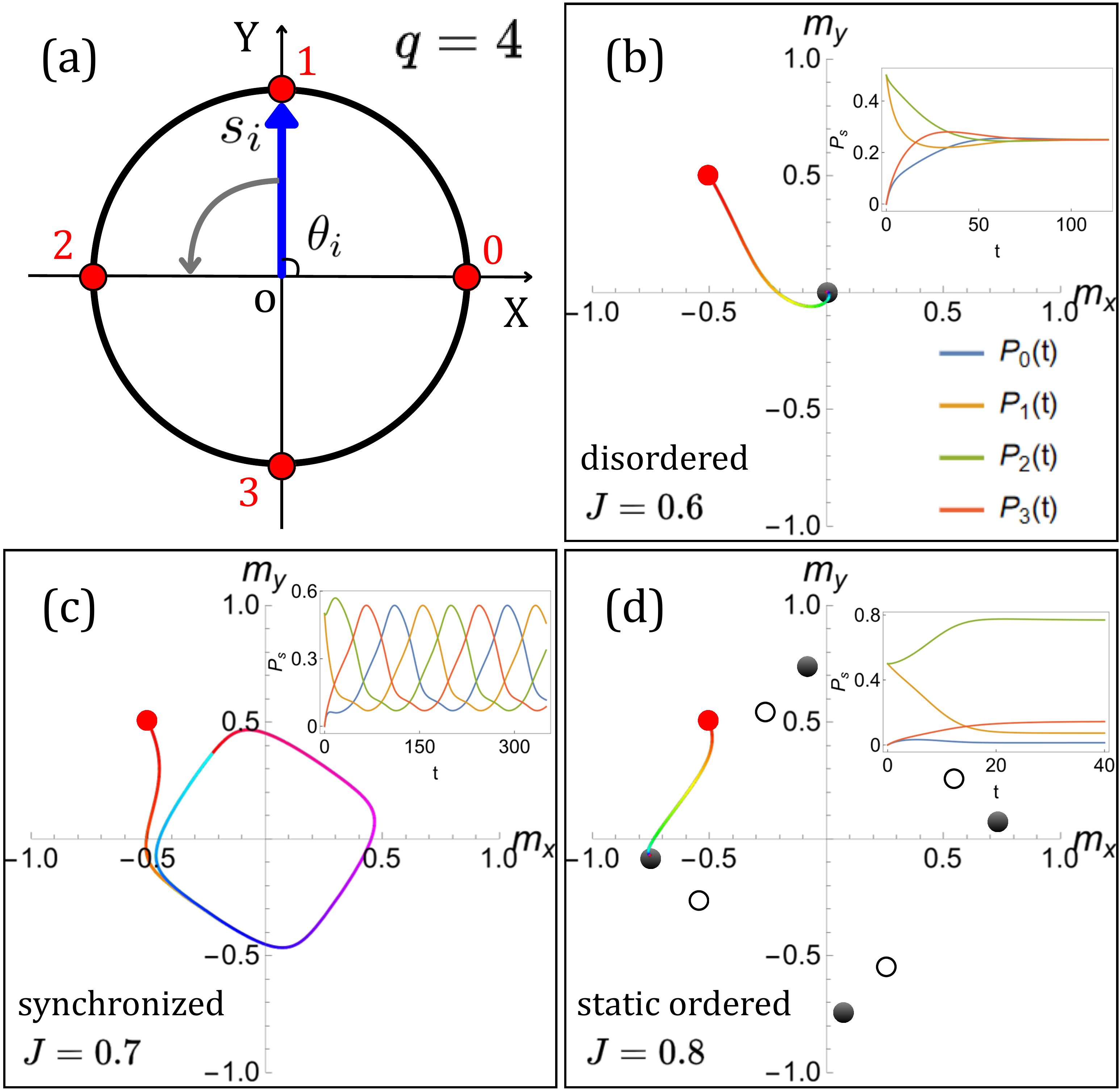}
    \caption{(a) Schematic of the self-rotating clock. Spin vector (blue arrow) points only along $q$ equidistant orientations (red dots). (b)-(d) Typical system trajectories in the $(m_x,m_y)$ plane before and after bifurcations, with $q=q_c^{MF}=4$ and $f = 0.12$ for $J=0.6$ (disordered phase), 0.7 (synchronized phase) and 0.8 (static ordered phase). Results are obtained from numerical solutions of the MFE.
    Black points: stable fixed points; hollow points: saddle fixed points; red points: starting points of dynamical trajectories. Trajectory color variation indicates time progression. Insets show the corresponding temporal evolutions of spin probabilities $\{P_s\}$.}
    \label{Fig:MFtraject}
\end{figure}

\emph{Model.}---
The $q$-state clock model is a spin model on a 2D $L\times L$ square lattice. Each spin can occupy $q$ discrete orientations (states) in the XY plane, denoted by an integer $s_i \in \{0, 1, \dots, q-1\}$ on site $i\in \{1,\dots,L^2\}$. The corresponding angle is $\theta_i = (2\pi/q) s_i$, representing $q$ equally spaced orientations on a unit circle (Fig.~\ref{Fig:MFtraject}(a)). The system configuration is thus specified by the vector $\vec{s} = (s_1, s_2, \dots, s_{L^2})$. In the absence of an external field, the Hamiltonian adopts the same form as the classical XY model:
\begin{equation}\label{Hamiltonian}
    \begin{aligned}
        H(\vec{s})=-J \sum_{\langle i,j \rangle} \cos(\theta_i - \theta_j),
    \end{aligned}
\end{equation}
where $J > 0$ is the ferromagnetic coupling coefficient, and $\langle i,j \rangle$ denotes a sum over all nearest-neighbor pairs with periodic boundary conditions. We apply a constant self-driven torque $f_0>0$ (counterclockwise) to each spin. The system evolves via a continuous-time Markov process. Each spin can only flip to its two adjacent states ($s_i \to s_i \pm 1$) via a rotation $R_i^{\pm}$. The rate for this rotation follows the Glauber dynamics:
\begin{equation}\label{Glauber}
    \begin{aligned}
        w(\vec{s}\to R^{\pm }_{i}\vec{s})=\frac{1}{2\tau}\left[1 - \tanh\left(\frac{\Delta E(\vec{s}\to R^{\pm }_{i}\vec{s})\mp f}{2k_B T}\right)\right],
    \end{aligned}
\end{equation}
with $T$ the temperature, $k_B$ the Boltzmann constant, and $\tau$ the characteristic timescale for a single rotation. Here, $\Delta E$ is the change in Hamiltonian \eqref{Hamiltonian} due to the spin rotation, and $f\equiv f_0 (2\pi/q) $ is the related work done by the self-driven torque. The dynamics satisfies local detailed balance but breaks global detailed balance (see SM for details). Unless otherwise specified, we set $k_B T = 1$ and $\tau = 1$ throughout.

\emph{Mean-field theory.}---
Before implementing accurate simulations, preliminary insights can be gained from mean-field theoretical calculations. From the cluster mean-field approximation and a spatially homogeneous ansatz, we obtain a $\mathbb{C}_q $-symmetric closed set of mean-field equations (MFE) governing the dynamics of single-spin probability distribution $P_s(t)$ (see SM for details). Numerical solution of the MFE for different $q$ reveals a critical value $q_c^{MF}=4$. For $q\ge q_c^{MF}$, a synchronized phase emerges. To visualize the high-dimensional dynamics and its bifurcations, we project $\{P_s\}$ onto the 2D mean spin plane as follows: $(m_x,m_y)=\sum_{s=0}^{q-1}P_s(\cos\frac{2\pi }{q} s,\sin\frac{2\pi }{q} s)$.

We fix the self-driven torque $f_0>0$ and increase the coupling strength $J$. For $q\ge q_c^{MF}$, the system undergoes two sequential bifurcations. First, a Hopf bifurcation destabilizes the origin fixed point on the $(m_x,m_y)$ plane, producing a $\mathbb{C}_q $-symmetric stable limit cycle (disordered-to-synchronized transition, Figs.~\ref{Fig:MFtraject}(b)(c) and Figs.~\ref{Fig:MFq35}(d)(e) in End Matter). Next, an infinite-period (or saddle node on an invariant circle) bifurcation replaces the limit cycle with $q$ pairs of $\mathbb{C}_q $-symmetric stable and saddle fixed points (synchronized-to-static ordered transition, Figs.~\ref{Fig:MFtraject}(c)(d) and Figs.~\ref{Fig:MFq35}(e)(f) in End Matter). For $q<q_c^{MF}$, the system undergoes a subcritical bifurcation with $\mathbb{C}_q$ symmetry (Figs.~\ref{Fig:MFq35}(a)-(c) in End Matter): it starts with one stable origin fixed point, and then $q$ $\mathbb{C}_q $-symmetric stable fixed points form and coexist with the origin. The origin destabilizes next, leaving only the $q$ non-zero fixed points—corresponding to a discontinuous disordered-to-static ordered phase transition.

\emph{Monte Carlo simulations.}---
To go beyond mean-field theory, we perform large-scale Monte Carlo (MC) simulations of 2D self-rotating clock models for $q=3,4,5,6$. At each simulation step, a spin is randomly selected, with a random trial rotation direction. The trial rotation is then accepted with a probability given by Eq.~\eqref{Glauber}. We define a sequence of $2\times L^2$ such updates as one MC step. Unless otherwise stated, the system is initialized to a fully ordered state (all spins aligned), and the results are accumulated in the steady state.

To distinguish different possible phases, we employ two complementary order parameters derived from the complex total magnetization
\begin{equation}\label{micromagnetic}
    \begin{aligned}
        Me^{i\Theta}=M_x+iM_y=\frac{1}{L^2} \textstyle \sum_{j=1}^{L^2} e^{i\theta_j},
    \end{aligned}
\end{equation}
where $M_{x,y}$ are the magnetization in X,Y direction. The first order parameter is
\begin{equation}\label{syn_odpa}
    \begin{aligned}
        R\equiv\left\langle M \right\rangle=\left\langle \sqrt{M_x^2+M_y^2} \right\rangle,
    \end{aligned}
\end{equation}
which quantifies the degree of spin alignment~\cite{Kuramoto}. Here, $\left\langle \dots \right\rangle$ denotes an average over time and realizations. The second is the angular momentum in the magnetization space~\cite{Nonreciprocal_pre}
\begin{equation}\label{angular_momentum}
    \begin{aligned}
        \mathcal{L} \equiv\left\langle M^2\partial_t \Theta \right\rangle=\left\langle M_x \partial_t M_y- M_y \partial_t M_x\right\rangle,
    \end{aligned}
\end{equation}
which measures the magnetization rotation rate and quantifies the macroscopic time-reversal symmetry breaking. 

\begin{figure}[htbp!]
    \centering
    \includegraphics[keepaspectratio, width=\columnwidth]{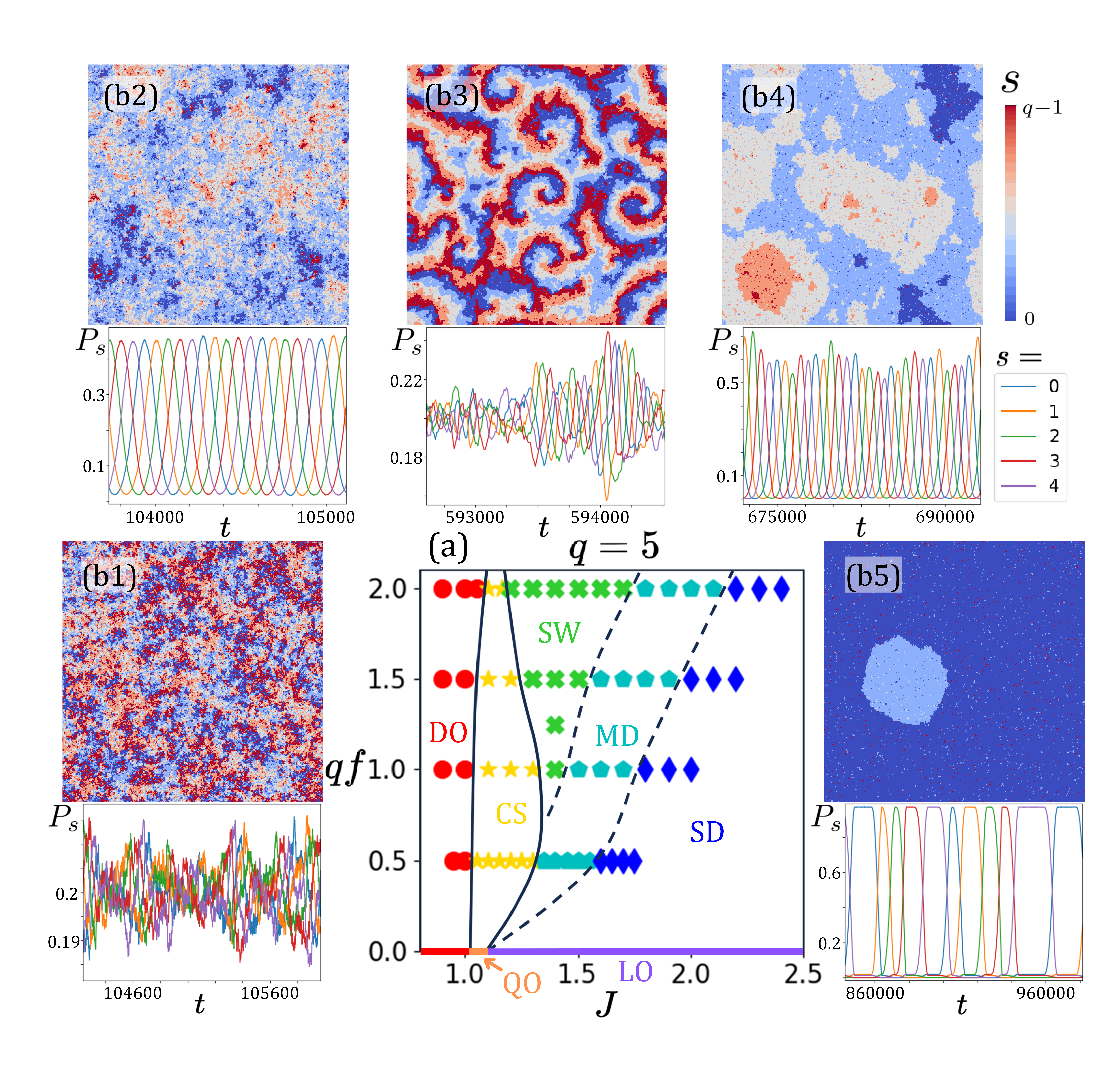}
    \caption{(a) Phase diagram in $(J, qf)$ plane for $q=q_c=5$, obtained from simulations with the system size $L=400$. Different symbols represent distinct phases. The solid lines are phase boundaries between the synchronized and other phases, while the dotted lines lie within the spatiotemporal pattern phase. Along $J$-axis, the equilibrium quasi-long-range ordered phase (QO) and long-range ordered phase (LO) are also marked. (b1)-(b5) show typical configuration snapshots (different colors represent different spin states) and corresponding temporal evolutions of spin percentage $P_s$, respectively, for disordered phase (DO, $f=0.1,J=0.9$), critical synchronized phase (CS, $f=0.1,J=1.1$), spiral wave state (SW, $f=0.4,J=1.6$), multi-droplet state (MD, $f=0.1,J=1.5$) and single-droplet state (SD, $f=0.1,J=1.7$).}
    \label{Fig:q5_pd}
\end{figure}
\begin{figure*}[ht]
    \centering
    \includegraphics[keepaspectratio, width=2\columnwidth]{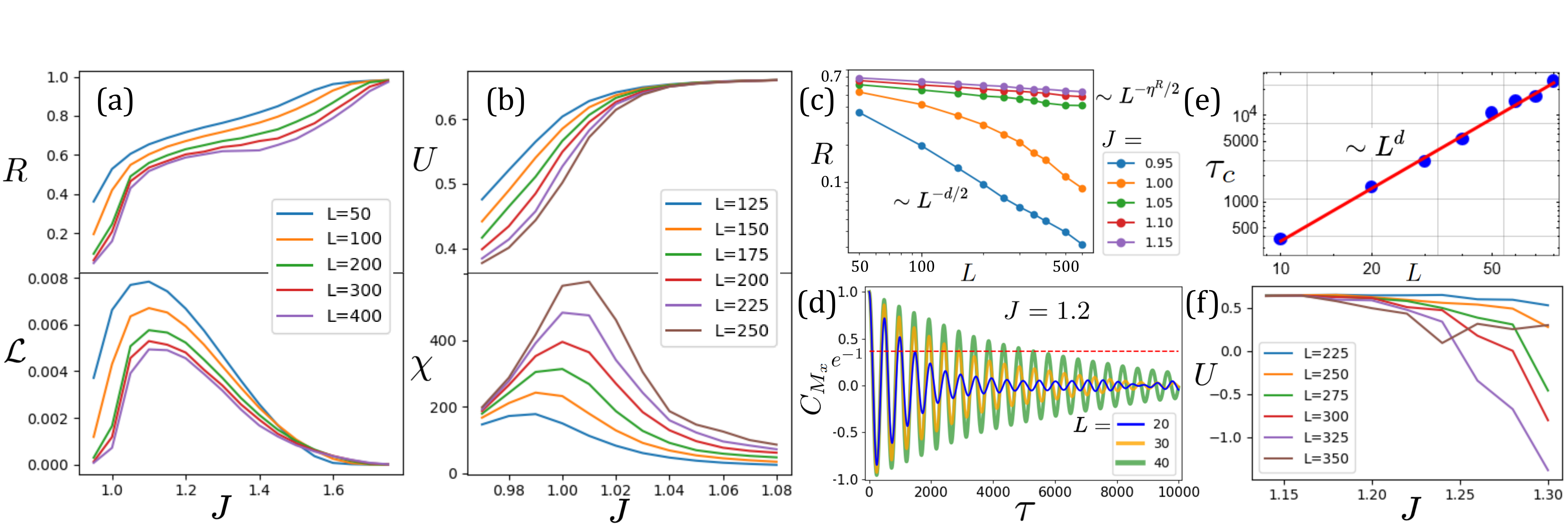}
    \caption{Critical synchronized phase and BKT-like transitions. (a) Variation of $R$ and $\mathcal{L}$ with $J$ for different system sizes $L$. (b) Variation of $U$ and $\chi$ with $J$ around critical point $J_{1}$. (c) Scaling relation of $R$ versus $L$ on a log-log plot for different $J$. (d) The autocorrelation function $C_{M_{x}}(\tau)$ of synchronized phase for different $L$. (e) The fitted scaling relationship between coherence time $\tau_c$ and $L$. In (a)-(e) $q=5$ and $f=0.1$ remain fixed. (f) $U$ around critical point $J_{2}$ at fixed $q=5$ and $f =0.4$, obtained from disordered initial configurations. For intermediate $L$ and larger $J$, $U < 1/3$ stems from the heavy-tailed probability distribution $P(M)$, which is induced by the intermittent annihilation of spiral waves (see SM for details).} 
    \label{Fig:KT}
\end{figure*}

\emph{Emergence of a 2D critical synchronized phase ($q\ge 5$).}---
Our large-scale simulations reveal that the mean-field prediction $q_c^{MF}=4$ is incorrect, but the true critical value is $q_c=5$. The system with $q= 5$ exhibits a variety of phases, including disordered, critical synchronized, spiral wave, multi-droplet and single-droplet states, as shown in Fig.~\ref{Fig:q5_pd}. The non-zero order parameters $R$ and $\mathcal{L}$ in Fig.~\ref{Fig:KT}(a) indeed signify the existence of a synchronized phase, where continuous U(1) symmetry emerges in the $\mathbb{C}_q$-symmetric $q$-state clock model (see SM for details), reminiscent of the quasi-long-range ordered phase in equilibrium 2D clock models for $q\ge 5$~\cite{einhorn_physical_1980,PhysRevE.111.054125}.

To determine the nature of the 2D phase transition between disordered and synchronized phases, we measure the Binder cumulant $U = 1 - \langle M^4 \rangle/3 \langle M^2 \rangle^2$~\cite{binder_finite_1981}, susceptibility $\chi = L^\mathrm{d} \left( \langle M^2 \rangle - \langle M \rangle^2 \right)$ ($\mathrm{d} = 2$ is the spatial dimension), and $R$ across different system sizes $L$ (Figs.~\ref{Fig:KT}(b)(c)). The curves $U$ for different $L$ merge into a single curve near the critical point $J_{1}(f=0.1)\approx1.05$—a hallmark of BKT-like phase transitions~\cite{loison_binders_1999,hasenbusch_binder_2008,clock_2022}. At $J=J_{1}$, the order parameter exhibits a scaling relation $R\sim L^{-\eta^R_{1}/2}$, with $\eta^R_{1}=0.319\pm0.004$, indicating that the spatial correlation begins to decay algebraically (since $R\sim L^{-\mathrm{d}}\sqrt{\sum_{i\ne j} \langle \cos(\theta_i-\theta_j) \rangle+L^{\mathrm{d}}}$, it implies $\langle \cos(\theta_i-\theta_j) \rangle \sim x^{-\eta^R_{1}}$, with $x$ the distance between sites $i$ and $j$). For $J > J_1$, the scaling $R\sim L^{-\eta^R/2}$ persists with an exponent $\eta^{R}(J)< \eta^{R}_1$, which decreases with $J$, confirming a critical synchronized phase. For $J < J_1$, we find $R\sim L^{-\mathrm{d}/2}$ (for example $R(J=0.97)\sim L^{-1.007\pm0.006}$), and the susceptibility approaches saturation with increasing size, characteristic of a disordered phase with an exponentially decaying spatial correlation. These findings demonstrate that as $J$ increases, the system undergoes a ``high-temperature" BKT-like transition from disordered to critical synchronized phases.

Besides synchronous oscillation and quasi-long-range order, the 2D synchronized phase exhibits another crucial property: its coherence time diverges with the system size, signaling the emergence of a time crystal~\cite{sacha_time_2017,Colloquium_crystals,Classical_Crystals,yao_classical_2020,Observation_time,wu_dissipative_2024}. To quantify this, we compute the autocorrelation function of the magnetization components:
\begin{equation}\label{autocorrelation}
    \begin{aligned}
        C_{M_{x,y}}(\tau) = \frac{\langle M_{x,y}(t)M_{x,y}(t+\tau) \rangle_t}{\langle M_{x,y}(t)^2 \rangle_t}.
    \end{aligned}
\end{equation}
Note that $\langle M_{x,y}\rangle_t=0$. Fig.~\ref{Fig:KT}(d) displays $C_{M_{x}}(\tau)$ of the synchronized phase for different system sizes $L$. The coherence time $\tau_c$ is defined as the time when the envelope of $C_{M_{x}}(\tau)$ decays to $e^{-1}$. A key finding is a power-law divergence of $\tau_c$ with $L$, with a fitted exponent $2.02\pm0.06$ (Fig.~\ref{Fig:KT}(e)). This $\tau_c(L)\sim L^\mathrm{d}$ scaling indicates that the synchronized phase of the 2D self-rotating clock model behaves as a continuous time crystal within observation scales~\cite{oberreiter_stochastic_2021,Nonreciprocal_pre}, exhibiting long-range temporal order through persistent coherent oscillations and breaking the continuous time translation symmetry.

\begin{figure}[htbp!]
    \centering
    \includegraphics[keepaspectratio, width=\columnwidth]{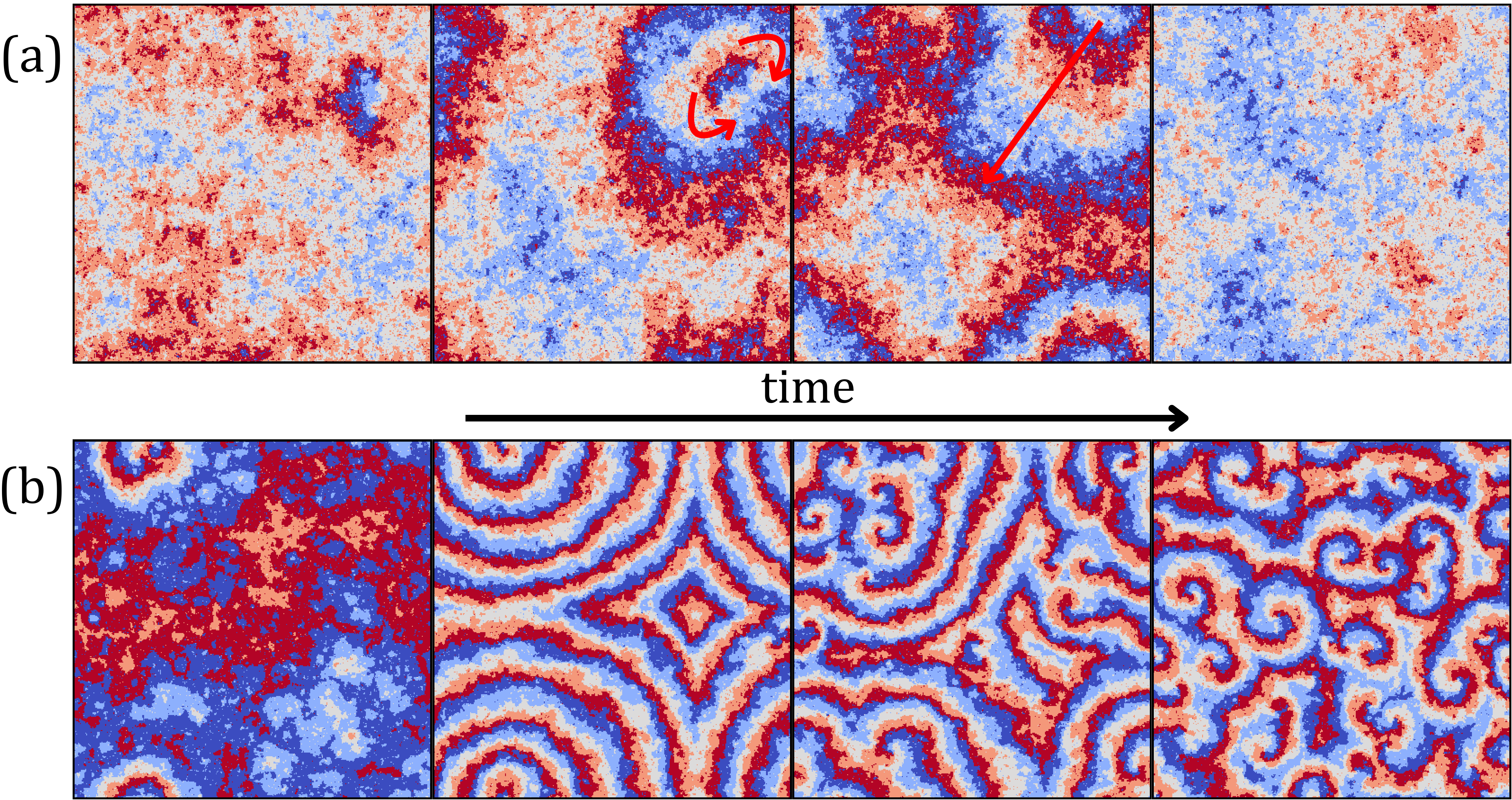}
    \caption{(a) Rapid annihilation of spiral cores after pairwise excitation in critical synchronized phase ($q=5, f=0.2, J=1.25,L=400$). The red arrows indicate the rotation direction of spiral cores and the propagation direction of spherical waves. (b) Formation of spiral wave state ($q = 5,f = 0.5,J = 1.7,L = 400$): A pair of stochastically excited spiral cores ultimately develops into spiral turbulence, disrupting global phase synchronization.}
    \label{Fig:spiral}
\end{figure}

Similar to the equilibrium 2D quasi-long-range ordered phase~\cite{kosterlitz_ordering_1973,einhorn_physical_1980}, the critical synchronized phase also features bound topological defect pairs: oppositely rotating vortex-antivortex pairs, manifesting as spiral core pairs. A defect pair is stochastically excited at nearby locations. After a finite lifetime, they randomly annihilate, forming 2D spherical or plane waves that widen and eventually dissipate (Fig.~\ref{Fig:spiral}(a)). Due to self-rotation and wave propagation, spiral core pairs disrupt global order more strongly than equilibrium vortex pairs ($\eta_{1}^{R}(f=0.1)>\eta_{1}^{R}(f=0)\approx0.234$~\cite{clock_2019}). In both equilibrium and non-equilibrium high-temperature BKT phase transitions, vortex pairs unbind, due to thermal fluctuations.

Because of self-rotation, the static ordered phase becomes unstable and develops into a spatiotemporal pattern phase, which encompasses three dynamic modes: multi-droplet, single-droplet and spiral wave states. In the thermodynamic limit ($L\to \infty$), only spiral wave state remains stable (see SM for details). Between critical synchronized phase and the pattern phase, we identify a ``low-temperature" BKT-like phase transition: the Binder cumulants for different system sizes merge below the critical point $J_{2}(f=0.4)\approx1.16$ (Fig.~\ref{Fig:KT}(f)). In the spatiotemporal pattern phase, domain walls—the topological defects of discrete-state systems—prevail dynamically at large scales. Their rigidity, combined with self-rotation, underpins the formation of large-scale $\mathbb{C}_q$-symmetric spiral waves (see SM for details). Upon transitioning to critical synchronized phase, the domain walls become floppy, condense and readily dissipate, consistent with the low-temperature BKT phase transition of equilibrium clock model~\cite{einhorn_physical_1980,ortiz_dualities_2012}. Unlike the synchronized phase, spiral core pairs in the spiral wave state can persist for a relative long period. Due to thermal perturbations and intrinsic wave instabilities~\cite{Doppler,Long-Wavelength}, wavefronts of a spiral wave (or the spherical wave emitted by a spiral core pair) can stretch, distort, and potentially breakup to generate new spiral tips (cores), that meanwhile undergo random meandering~\cite{meandering,diks_spiral_1995}. Once initiated, this process develops as a chain reaction. Eventually, the system evolves into a spatiotemporal chaotic state filled with numerous spirals that continuously generate and annihilate (Fig.~\ref{Fig:spiral}(b)), analogous to spiral turbulence in excitable and oscillatory media~\cite{bar_turbulence_1993,Chemical_Turbulence}. These observations imply that any finite self-driven torque can make the unstable vortices in equilibrium long-range ordered phase become thermodynamically stable (also see SM).

\emph{$q$-dependent phase diagrams.}---
For $q < 5$, the system exhibits only two phases: disordered and spatiotemporal pattern phases, with the latter containing spiral wave and single-droplet states in large but finite systems (see SM for details). In contrast, for $q > 5$, an intermediate critical synchronized phase emerges, and the pattern phase also includes a multi-droplet state, as the case of $q=5$ (Fig.~\ref{Fig:q5_pd}(a)). The dependence of phase behavior on $q$ is parallel to equilibrium clock models. Complete phase diagrams of the 2D self-rotating clock models for $q=3,4,6$ and $\infty$ can be found in End Matter.

Notably, when $q$ approaches infinity with $qf$ held constant, our model naturally transitions to the Kuramoto model composed of locally-coupled identical oscillators under Gaussian white noise~\cite{sarkar_noiseinduced_2020,Self_Driven_Rotors}. A uniform rotating-frame transformation then maps the system onto the equilibrium XY model, thereby demonstrating that its phase diagram is independent of self-driven torques. This limit behavior is reflected in the following trend: as $q$ increases, the high-temperature BKT-like transition line gradually becomes more parallel to the $qf$-axis, while the low-temperature BKT-like transition line extends gradually to larger $J$. In the $q\to \infty$ limit, the high-temperature BKT phase transition line is independent of $qf$, and the spatiotemporal pattern phase vanishes.

\emph{Renormalization group analysis.}---
To investigate the stability of critical synchronized phase in the thermodynamic limit and elucidate the $q$-dependent phase behavior, we construct a coarse-grained field theory for the local order parameter $\psi = m e^{i\theta}$ with $0 \le m \le 1$ (see SM for details). Deep in the synchronized phase, the amplitude $m$ quickly relaxes to its steady-state value $m_0$ (determined by the Ginzburg-Landau free energy), and the phase dynamics reduces to
\begin{equation}\label{p_pde}
    \begin{aligned}
        \partial_t \theta = \nu \nabla^2 \theta - u \sin(q\theta) + f_d + \xi,
    \end{aligned}
\end{equation}
where $\nu$, $f_d$ and $u\sim m_0^{q-2}$ are phenomenological coefficients and $\xi$ is the white noise with strength $D$. A dynamic renormalization group (RG) analysis treats the clock potential term $- u \sin(q\theta)$ perturbatively (see SM for details). At first order, thermal fluctuations coarse-grain the discrete clock potential. For $q \ge 5$ and $\frac{2}{\pi} < \frac{\nu}{D} < \frac{q^2}{8\pi}$, this makes $u$ irrelevant under RG, leading to an emergent U(1) symmetry and 2D Edwards-Wilkinson (EW)-type quasi-long-range order—the critical synchronized phase. However, at second order, the combination of clock potential and self-rotation generates a KPZ nonlinearity $\frac{\lambda}{2} (\nabla \theta)^2$. While its coefficient saturates at a small value $\lambda_\infty\sim \omega_0u^2$ for macroscopic systems, this KPZ term dictates that true quasi-long-range order must be destroyed as $L \to \infty$~\cite{grinstein_temporally_1993,daviet_kardar-parisi-zhang_2025,fruchart_nonreciprocal_2026}. The crossover scale from EW to KPZ universality, whose general formulation follows Ref.~\cite{nattermann_kinetic_1992}, grows double-exponentially with $q$:
\begin{equation}\label{KPZ_crossover}
    \begin{aligned}
        L^* \sim \exp\left( \nu^3/\lambda_\infty^2 D \right) \sim \exp\left( \omega_0^{-2} m_0^{-4q} \right),
    \end{aligned}
\end{equation}
where $\omega_0$ is the macroscopic rotation rate. For $q \ge 5$, $L^*$ can vastly transcend typical simulation dimensions, ensuring the robust pre-asymptotic stability of the synchronized phase. 
Furthermore, by measuring the collective rotation rate for different winding numbers~\cite{manneville_phase_1996}, we fit the KPZ coefficient and estimate that deep in the synchronized phase $L^*$ readily exceeds $10^8$ lattice spacings (see SM for details). Beyond $L^*$, the system enters phase turbulence induced by KPZ nonlinearity. While this continuum KPZ turbulence and the discrete spiral wave state feature distinct underlying symmetries, they both characterize the spatiotemporal chaotic regime where synchronization is destroyed. For $q < 5$, the clock potential remains relevant, precluding the synchronized phase. This yields critical $q^{RG}_c =5$, in full agreement with our simulation results and initial hypothesis.

\emph{Conclusion.}---
Our work provides the first demonstration of a critical synchronized phase emerging on macroscopic scales in 2D discrete-state oscillator systems with short-range interactions and thermal fluctuations. Although asymptotically destroyed by KPZ nonlinearity, this synchronization remains robustly protected by a double-exponentially diverging crossover scale. On accessible scales, the synchronized phase manifests as a continuous time crystal bounded by two distinct BKT-like transitions into disordered and spatiotemporal pattern regimes.
These findings establish the minimal self-rotating clock model as a versatile platform for exploring nonequilibrium phase behavior and cooperative phenomena across diverse fields, including chemistry~\cite{mikhailov_control_2006,novak_design_2008}, biology~\cite{beta_intracellular_2017,bailles_mechanochemical_2022}, and ecology~\cite{kerr_local_2002,moura_behavioural_2021}.

\vspace{1em}
\emph{Acknowledgments.}---
We thank Hugues Chaté, Longfei Li, Jonas Köppl and Xia-qing Shi for helpful discussions. This work was supported by the National Key R\&D Program of China (2022YFF0503504) and National Natural Science Foundation of China (No. T2325027, 12274448).

\bibliographystyle{apsrev}
\bibliography{RefList}
\clearpage
\onecolumngrid

\section*{End Matter}

\twocolumngrid

\renewcommand\thefigure{A\arabic{figure}}  
\renewcommand\thetable{A\arabic{table}}  
\renewcommand{\theequation}{A\arabic{equation}}
\setcounter{equation}{0}
\setcounter{figure}{0}
\setcounter{table}{0}
\emph{Appendix A: Results of mean-field theory for $q=3,5$}
\vspace*{1pt}
\begin{figure}[H]
    \centering
    \includegraphics[keepaspectratio, width=\columnwidth]{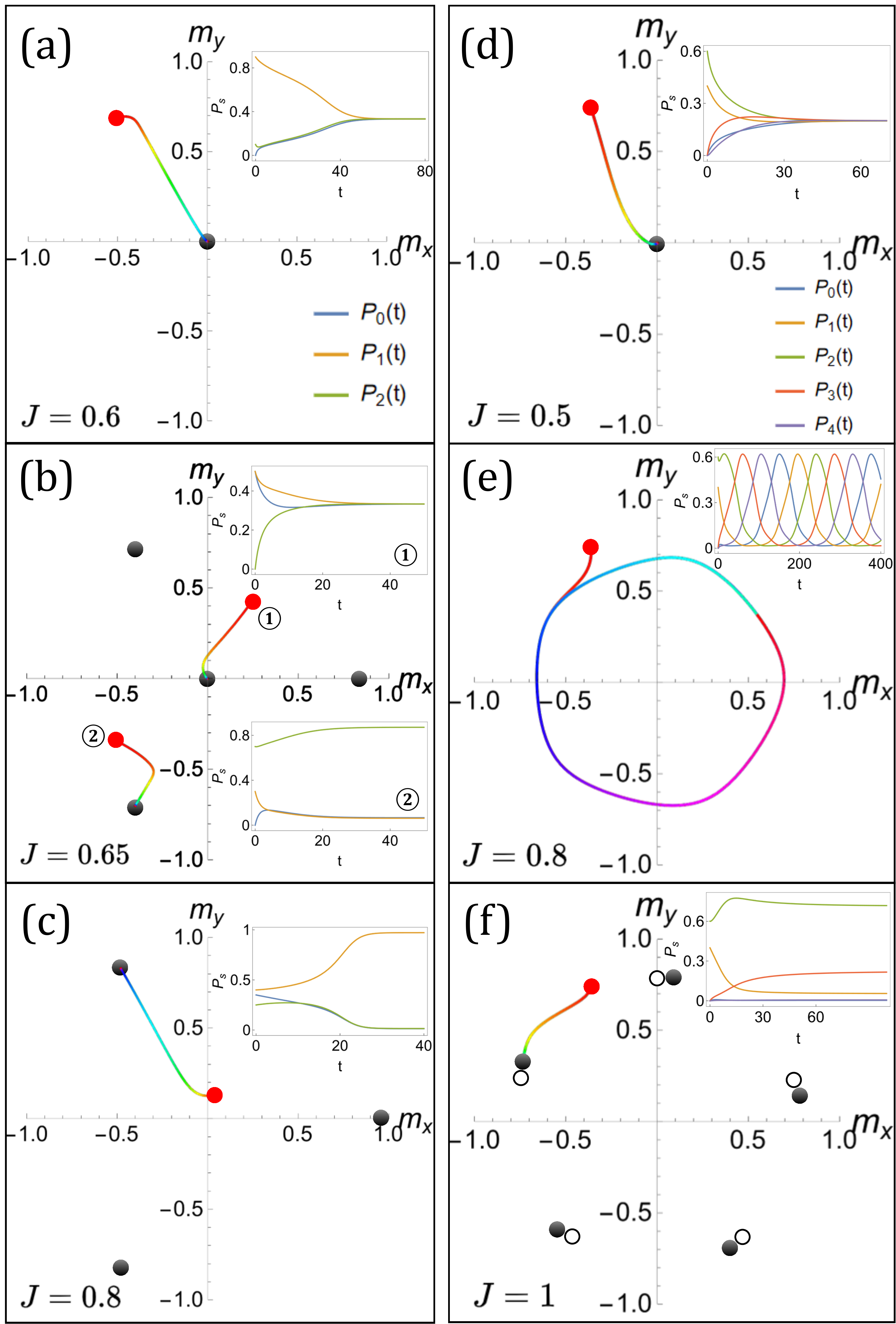}
    \caption{Projections of typical system trajectories in the $(m_x,m_y)$ plane before and after bifurcation, for different $q$ values at fixed $qf = 0.48$. (a)-(c) $q = 3$ and $J = 0.6, 0.65, 0.8$ (disordered, coexisting and static ordered phases). (d)-(f) $q=5$ and $J=0.5, 0.8, 1$ (disordered, synchronized and static ordered phases).}
    \label{Fig:MFq35}
\end{figure}

\renewcommand\thefigure{B\arabic{figure}}  
\renewcommand\thetable{B\arabic{table}}  
\renewcommand{\theequation}{B\arabic{equation}}
\setcounter{equation}{0}
\setcounter{figure}{0}
\setcounter{table}{0}
\emph{Appendix B: Phase diagrams for $q=3,4,6$ and $\infty$}
\vspace*{3pt}
\begin{figure}[H]
    \centering
    \includegraphics[keepaspectratio, width=\columnwidth]{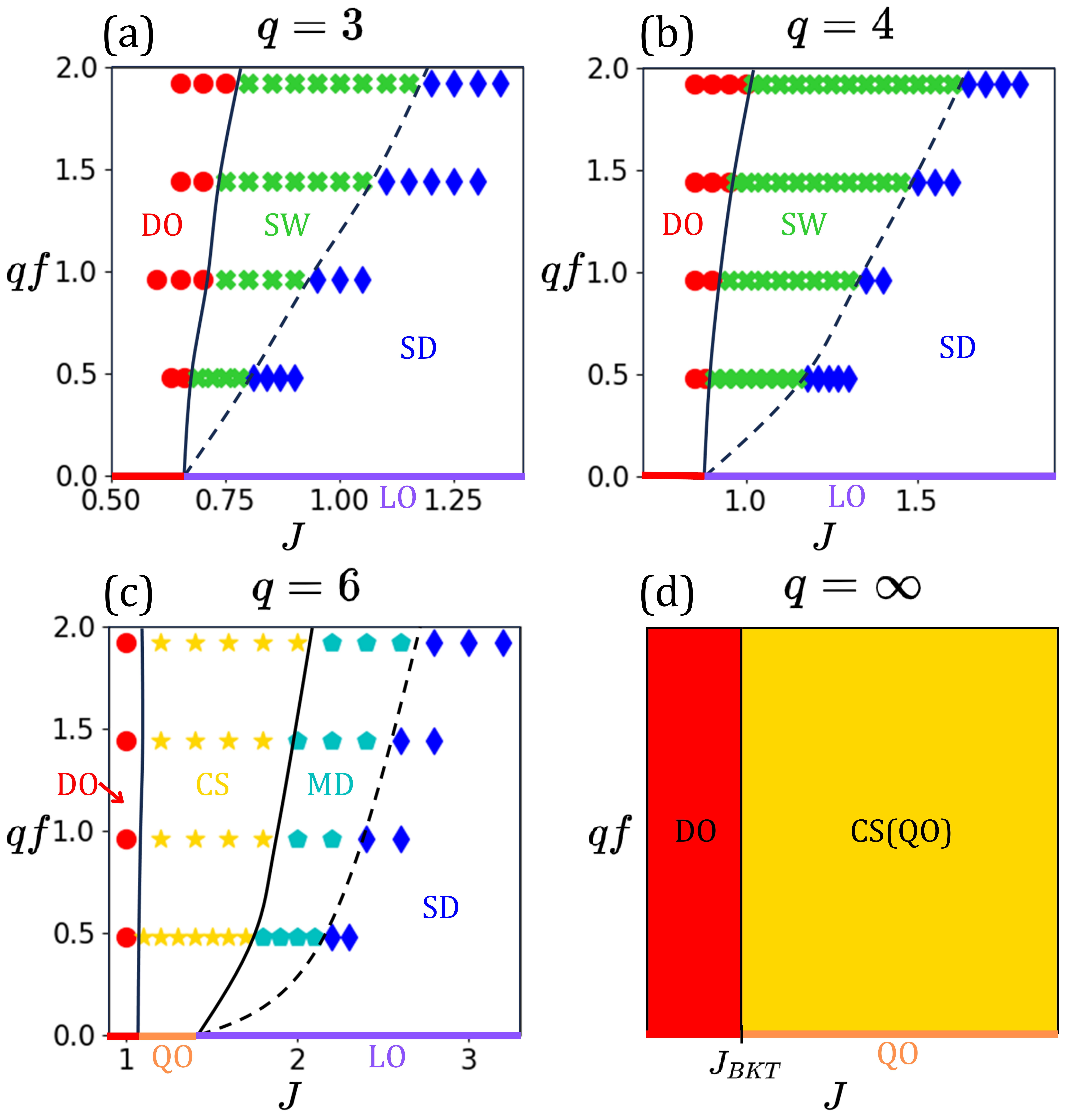}
    \caption{$q$-dependent phase diagrams in $(J, qf)$ plane. For $q < 5$, the system exhibits only disordered phase (DO), spiral wave state (SW), and single-droplet state (SD). For $q > 5$, critical synchronized phase (CS) and multi-droplet state (MD) emerge. (a)-(c) depict phase diagrams determined from MC simulations with the system size $L=400$ at $q=3, 4$ and 6, respectively. (d) shows the theoretically predicted phase diagram at $q=\infty$, where $J_{BKT}$ is the reduced coupling strength for the BKT phase transition in equilibrium 2D XY model.}
    \label{Fig:phase}
\end{figure}

\end{document}